\documentclass[%
 reprint,
 superscriptaddress,
 showpacs, 
 amsmath,amssymb,
 aps,
 pre,
 floatfix,
]{revtex4-1}

\usepackage[pdftex]{graphicx}
    \DeclareGraphicsExtensions{.pdf,.png}
\usepackage{dcolumn}
\usepackage{bm}
\usepackage{xcolor}

\begin{document}

\title{Self-diffusion in two-dimensional quasi-magnetized rotating dusty plasmas}

\author{P.~Hartmann}
\affiliation{Institute for Solid State Physics and Optics, Wigner Research Centre, Hungarian Academy of Sciences, P.O. Box 49, H-1525 Budapest, Hungary}
\affiliation{Center for Astrophysics, Space Physics and Engineering Research (CASPER), One Bear Place 97310, Baylor University, Waco, Texas 76798, USA}

\author{J.~C.~Reyes}
\affiliation{Center for Astrophysics, Space Physics and Engineering Research (CASPER), One Bear Place 97310, Baylor University, Waco, Texas 76798, USA}

\author{E.~G.~Kostadinova}
\affiliation{Center for Astrophysics, Space Physics and Engineering Research (CASPER), One Bear Place 97310, Baylor University, Waco, Texas 76798, USA}

\author{L.~S.~Matthews}
\affiliation{Center for Astrophysics, Space Physics and Engineering Research (CASPER), One Bear Place 97310, Baylor University, Waco, Texas 76798, USA}

\author{T.~W.~Hyde}
\affiliation{Center for Astrophysics, Space Physics and Engineering Research (CASPER), One Bear Place 97310, Baylor University, Waco, Texas 76798, USA}

\author{R.~U.~Masheyeva}
\affiliation{IETP, Al Farabi Kazakh National University, 71, al Farabi Avenue, Almaty, 050040, Kazakhstan}

\author{K.~N.~Dzhumagulova}
\affiliation{IETP, Al Farabi Kazakh National University, 71, al Farabi Avenue, Almaty, 050040, Kazakhstan}

\author{T.~S.~Ramazanov}
\affiliation{IETP, Al Farabi Kazakh National University, 71, al Farabi Avenue, Almaty, 050040, Kazakhstan}

\author{T.~Ott}
\affiliation{Institute for Theoretical Physics and Astrophysics, Christian-Albrechts-University Kiel, Leibnizstrasse 15, 24098 Kiel, Germany}

\author{H.~K\"ahlert}
\affiliation{Institute for Theoretical Physics and Astrophysics, Christian-Albrechts-University Kiel, Leibnizstrasse 15, 24098 Kiel, Germany}

\author{M.~Bonitz}
\affiliation{Institute for Theoretical Physics and Astrophysics, Christian-Albrechts-University Kiel, Leibnizstrasse 15, 24098 Kiel, Germany}

\author{I.~Korolov}
\affiliation{Institute for Solid State Physics and Optics, Wigner Research Centre, Hungarian Academy of Sciences, P.O. Box 49, H-1525 Budapest, Hungary}

\author{Z.~Donk\'o}
\affiliation{Institute for Solid State Physics and Optics, Wigner Research Centre, Hungarian Academy of Sciences, P.O. Box 49, H-1525 Budapest, Hungary}

\date{\today}

\begin{abstract}
The self-diffusion phenomenon in a two-dimensional dusty plasma at extremely strong (effective) magnetic fields is studied experimentally and by means of molecular dynamics simulations. In the experiment the high magnetic field is introduced by rotating the particle cloud and observing the particle trajectories in a co-rotating frame, which allows reaching effective magnetic fields up to 3000 Tesla. The experimental results confirm the predictions of the simulations: (i) super-diffusive behavior is found at intermediate time-scales and (ii) the dependence of the self-diffusion coefficient on the magnetic field is well reproduced.
\end{abstract}

\maketitle

\section{\label{sec:intro} Introduction }

From the first observation of plasma crystals \cite{pk1,pk2,pk3,Hayashi94} two decades ago, strongly coupled dusty plasmas have been serving as a uniquely useful, simple, and universal ``tool'' for the studies of various physical processes in weakly damped, strongly coupled many-particle systems. The ensemble of electrically charged solid micro-particles that are levitated in a gas discharge plasma qualitatively reproduces most of the features of simple atomic matter, but at length- and time-scales that are easily accessible with standard video microscopy techniques providing direct visual access to individual particle trajectories. Utilizing this property, dusty plasmas can be used to investigate the microscopic details of classical macroscopic phenomena like collective excitations, thermal conductivity, viscosity, diffusion, deformation of crystalline solids, liquid flow including turbulence, phase transitions, etc., as well as phenomena like self-organization, lattice defect formation and migration, etc. Experiments on one-, and two-dimensional (2D) systems (particle chains and single layers) are performed routinely in ground based laboratories, but for three-dimensional systems microgravity environments provide obvious advantages \cite{Nefedov,PK3P,PK4}.

In most of the studies to date the observed motion of the dust grains could be described by theoretical models (accompanied by computer simulations) that de-coupled the dust dynamics from the complex interaction of the grains with the gas discharge plasma, which serves as the embedding medium and provides the means for the particles to acquire their charge. The basis of this reasonable approximation lies in the very different time-scales characterizing the discharge plasma and the charged dust ensemble. Both in radio frequency (RF) and direct current (DC) discharges the typical response times are in the nanosecond to microsecond regime for electrons and ions, while dust grains with typical diameters above 1 micron react on the scale of 10 to 100 milliseconds. Phenomena like charge fluctuation on the grain surface and the alternating external electric field in the RF case can be neglected: the grains experience only the time averaged effect of these. Assuming a homogeneous background discharge plasma, the most simple model that is still widely used and most successful for 2D systems is the Yukawa One Component Plasma (YOCP) model, which approximates the net interaction between the dust and the plasma via a single exponential screening parameter as derived by Debye and H\"uckel for electrolytes \cite{DebyeHuckel}. More realistic models (and, for example, Langevin dynamics simulations) take into account the driven-dissipative nature of the experiments and have successfully quantified the effect of the background gas \cite{langevin,Schweigert1998,Wang2018,Feng14}. The fact that in most cases the behavior of the dust grains that are part of a complex, multi-component system called dusty plasma can be described by the YOCP model is another essential reason why dusty plasmas are such a universal tool. 


The idea to extend strongly coupled dusty plasma research into the exciting world of magnetized systems emerged in the early years of experimental complex plasma research. Pioneering experimental work has been performed by U. Konopka \cite{Konopka00} and N. Sato \cite{Sato01} with promising results. These and later experiments \cite{Konopka05,Konopka09} inspired several groups to perform numerical simulations \cite{Uchida04,Bonitz10,OttDiffu} of magnetized dusty plasmas and provided the foundations for the new generation of magnetic dusty plasma experiments (MDPX) currently in operation \cite{Ed12}. During the interpretation of the experimental observations, two fundamental conclusions were reached that we interpret as problems that hinder the above mentioned universal applicability of dusty plasmas as simple model systems of atomic matter in magnetic fields. These circumstances are:

First, due to the small charge-to-mass ratio of the dust grains relative to that of electrons and ions, extremely large magnetic fields (in the range of thousands of Teslas) are needed to magnetize the dust component. Here magnetization is defined as a state where the cyclotron motion (radius and frequency) of the dust particles is comparable to that of the dust dynamics (inter-particle distance, plasma frequency of the dust particles).

Second, using a magnetic field strong enough to magnetize the dust greatly modifies the dynamics of the electrons and ions in the gas discharge and introduces practically unpredictable inhomogeneities and anisotropies of the electron and ion densities and fluxes \cite{Konopka05}. The dust particles are sensitive to these inhomogeneities and settle into structures that are defined by the background plasma, and not by the inter-grain interactions. This finding does on one hand open new interesting research directions \cite{Ed16}, but on the other hand makes the separation of the dust dynamics and discharge plasma dynamics impossible.

A possible solution that overcomes both issues was suggested by H. K\"ahlert et. al. \cite{Kahlert12,Bonitz13} based on the Larmor-theorem \cite{Larmor}. Using the formal equivalence of the magnetic Lorentz force $F_{\rm m} \sim Q\,{\bf v}\times{\bf B}$ and the Coriolis force $F_{\rm C} \sim 2m\,{\bf v}\times{\bf \Omega}$, one can be substituted for the other. Here $Q$ and $m$ are the electric charge and mass of the dust grain, the vectors ${\bf v}$, ${\bf B}$, ${\bf \Omega}$ are the velocity, magnetic induction, and the angular velocity of rotation of the (whole) system. Although the Coriolis force is not present in the laboratory reference frame, it appears as an inertial force together with the centrifugal force if one observes the rotating system from a co-rotating frame. In this case the particle velocities are defined in this co-rotating reference frame. The applicability of this idea was first demonstrated on the vibration spectrum of a small cluster of dust grains \cite{Kahlert12} in an experimental setup introduced in \cite{Carstensen10} and later on the collective excitation spectra and wave dispersion properties of a 2D many-particle ensemble \cite{RotoDust} in the ``RotoDust'' setup. It has been shown that this alternative approach solves both issues that arise in real magnetic dusty plasma experiments: the equivalent magnetic field can be extremely high, and the plasma properties (primarily the homogeneity, as well as the electron and ion dynamics) are practically unaffected by the rotation of one of the electrodes, due to the large difference in time-scales of the rotation (few Hz) and the electron and ion motions. In this way, the RotoDust experiment successfully extends the applicability of dusty plasmas to study principal many-body phenomena at the particle level in a universal fashion to magnetized systems.

In this article we focus on self-diffusion, one of the fundamental transport processes in nature,  in magnetized strongly coupled dusty plasmas in the liquid phase. RotoDust experiments were performed in the Hypervelocity Impacts and Dusty Plasmas Lab (HIDPL) of the Center for Astrophysics, Space Physics, and Engineering Research (CASPER) at Baylor University, Waco, Texas, and at the Institute for Solid State Physics and Optics, part of the Wigner Research Centre for Physics of the Hungarian Academy of Sciences, Budapest (referenced, respectively, as ``TEX'' and ``BUD'' in the following). Details of the dusty plasma apparatuses can be found in an earlier publication \cite{Harti14}, here only a brief outline is given in Section~\ref{sec:exp}, where details of the methods of the measurements and data evaluation are given. Molecular dynamics (MD) simulations, described in Section~\ref{sec:sim}, are also performed to compute the mean square displacement and diffusion coefficient of 2D magnetized Yukawa systems with systems parameters matching the experimental conditions for comparison. The results are presented in Section~\ref{sec:res}, while a summary is given in Section~\ref{sec:sum}.

\section{\label{sec:exp} Rotodust experiments and data evaluation}

\begin{figure}[htbp]
\begin{center}
\includegraphics[width=\columnwidth]{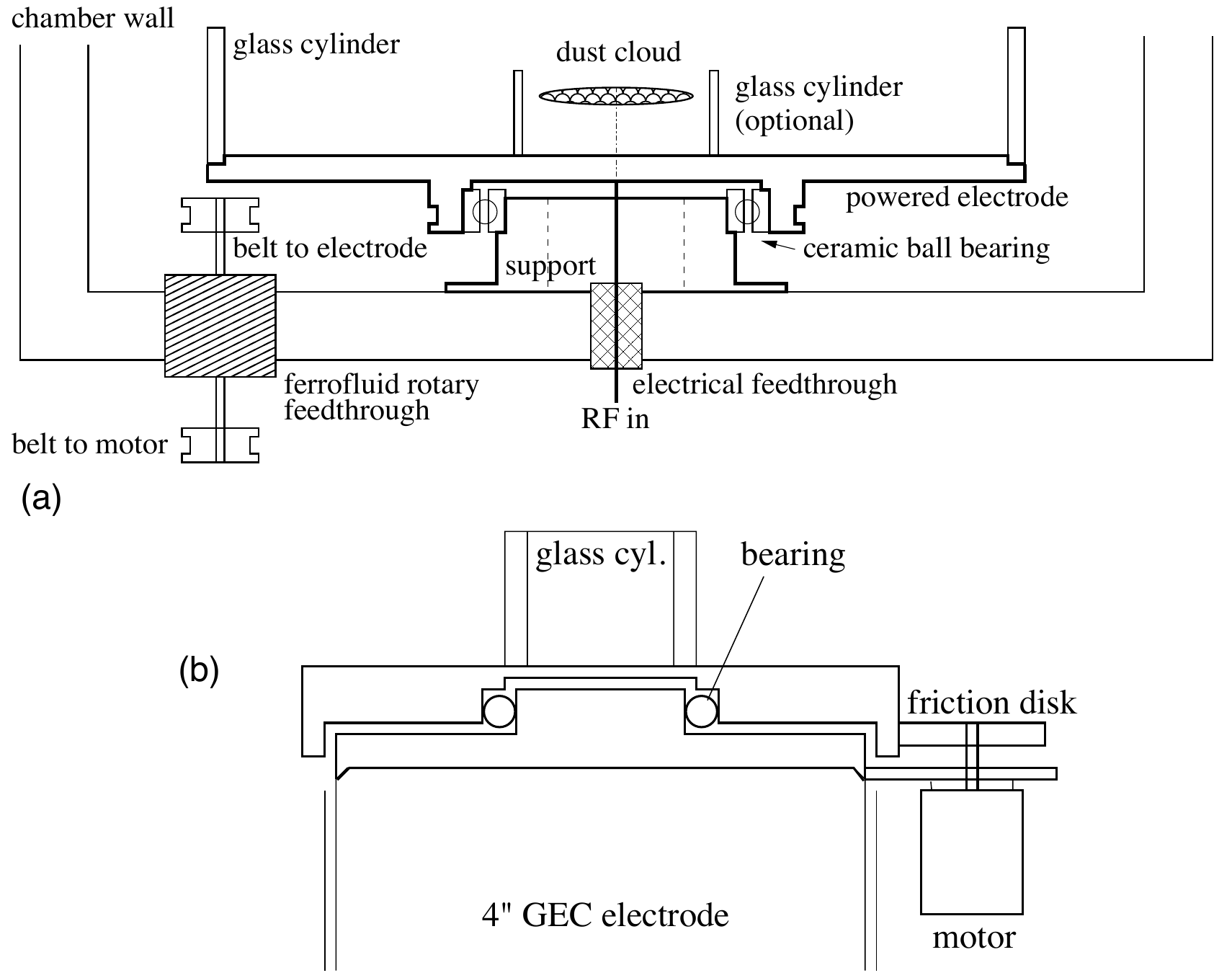}
\caption{Schematics of the ``RotoDust'' electrode configuration of the (a) BUD setup and the (b) TEX system.}
\label{fig:setup}
\end{center}
\end{figure}

Both the TEX and BUD experiments are based on RF discharges operated at frequency of 13.56 MHz in argon gas, with rotatable horizontal powered lower electrodes, as shown schematically in figure~\ref{fig:setup}. The relevant difference between the two setups is in the gas pressure operation regime due to the size difference of the respective plasma volumes. In the TEX experiments the discharge gap was 2.5~cm and the gas pressure was varied between 10 and 30 Pa, while in the BUD experiment the discharge gap is 15~cm and gas pressures between 0.5 and 1.5~Pa were used. 

In both setups the horizontal electrostatic confinement was enhanced by glass cylinders with inner diameters 1/4 to 1/2 inches placed on the rotating electrode, with a careful alignment of the symmetry axis of the glass cylinder and the axis of rotation. This enhancement is necessary to compensate for the centrifugal effect that acts against the horizontal confinement. The rotation of the lower electrodes are driven by controllable speed dc motors (of type BMU260C-A-3) through a ferrofluid rotary vacuum-feedthrough in the BUD system, or from inside the vacuum chamber in the TEX setup. The rotating electrode drags the gas inside the glass cylinder, transferring the rotation to the dust particles by neutral drag. Turbulent motion of the neutral gas is not expected, as the Reynolds number is very low in such a low-pressure environment. After turning on the motor, it takes about a second for the dust cloud to reach the rotation rate of the electrode.

Melamine-formaldehyde (MF) particles with diameters of $d_{\rm TEX} = 8.89 \,\mu{\rm m} \pm 1\%$ and $d_{\rm BUD} = 4.38 \,\mu{\rm m} \pm 1\%$ were dispersed into the glass cylinders while operating the discharge plasmas in the range of 2 to 20 Watts of RF power. The experimental parameters (RF power, gas pressure, and rotation rate) were adjusted to achieve large homogeneous single layer configurations in the strongly coupled liquid regime. Typical particle numbers in the dust clouds ranged from $N=100$ to 1000 grains with the diameter of the dust cloud ranging from 3 to 10\,mm. The dust clouds were illuminated with wide laser beams and image sequences were recorded at 125 frames per second (fps). The image exposure time was set to 1/2000 seconds to prevent the images of individual particles from streaking given the fast rotation in the range of 2 to 6 revolutions per second.

\begin{figure}[htb]
\begin{center}
\includegraphics[width=0.9\columnwidth]{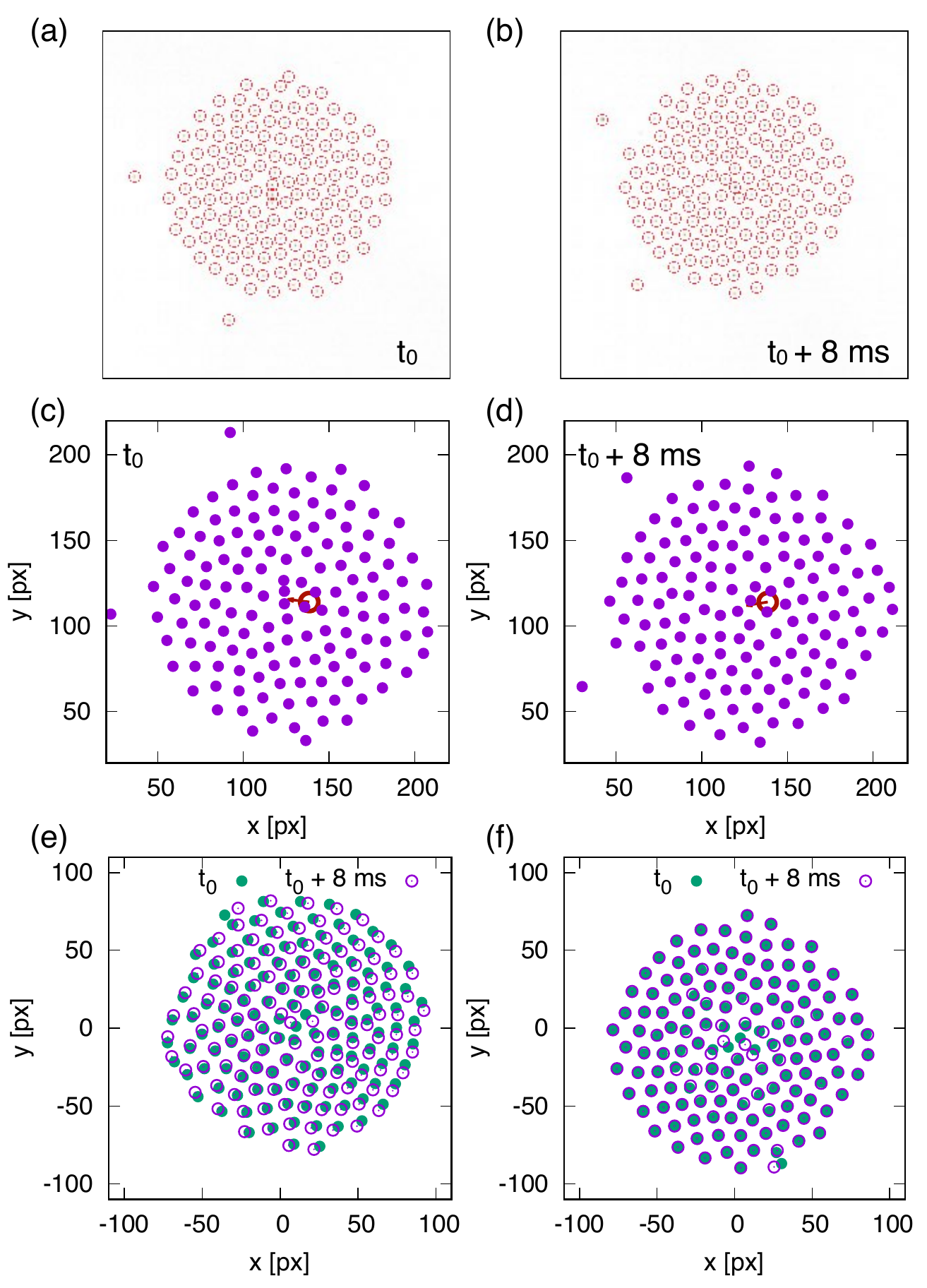}
\caption{Elementary steps of the de-rotation transformation demonstrated using two subsequent snapshots from experiment TEX11 (see Table ~\ref{tab:TEX}, with recording rate 125 frames per second, image resolution 256 x 256 pixels, and a field of view of 4.5 x 4.5 mm.). Raw images at times (a) $t_0$ and (b) $t_0$ + 8 ms, with superposed red circles showing the result of the particle detection. (c) and (d) Detected particles together with the center of rotation (COR, red circle) and the vector connecting the COR with the instantaneous center of mass (COM). (e) Overlay of the two sets of particle positions after the rough de-rotation described in Step 5. (f) Overlay of the two sets of particle positions after the least-square correction and removal of the outer particle ring relative to COR as described in Step 6.}
\label{fig:COM}
\end{center}
\end{figure}

The optical magnification is chosen to entirely fit the whole ensemble into the observation field-of-view. The image sequences were processed by our own algorithm, which includes the following steps:
\begin{enumerate}
\item Particle detection was performed following the procedure described in \cite{Feng07}. At this stage the apparent displacements of particles between subsequent images are too large for direct tracing, see fig.~\ref{fig:COM}(a,b).
\item The center of mass (COM) for each frame is calculated from the positions of all the particles in each frame.
\item Over time, the position of the COM is observed to move in a small circle. The center of rotation (COR) is identified as the long time average of the COM positions found for each frame.
\item An initial estimate of the rotation rate $\Omega$ is derived from the variation of the vector connecting the COR and COM from frame to frame, see fig.~\ref{fig:COM}(c,d).
\item Particle coordinates in the rotating frame are found by applying the inverse rotation with angular velocity $-\Omega$ about the point COR to compensate for the overall rotation, see fig.~\ref{fig:COM}(e).
\item A periodic artificial ``wobbling'' of the derived coordinates of the dust cloud due to small misalignment of the axis of rotation and the symmetry axis of the glass cylinder is compensated by applying a least-square minimization algorithm to the differences of particle positions in subsequent frames. Parameters found in this step are the additional frame-to-frame translation and rotation of the particle cloud that is superposed on the steady rotation already subtracted during steps 4 and 5, see fig.~\ref{fig:COM}(f).
\item To obtain the trajectories of individual particles, the grains have to be identified from frame to frame. This is done by linking the grain with the nearest position in the next frame to the particle in the current frame.   
\end{enumerate}

As a result of these data processing steps, one obtains the 2D coordinates and velocities of all particles as observed in a co-rotating frame as functions of time for sequences of typically 10\,000 to 40\,000 frames. Using these particle positions the mean square displacement
\begin{equation}
{\rm MSD} = \langle s^2 \rangle(\tau) = \left\langle\frac{1}{N} \sum_{i=1}^N \left[{\bf r}_i(t) - {\bf r}_i(t+\tau) \right] ^2 \right\rangle_t
\end{equation}
can be easily measured.

For ideal systems the diffusion coefficient can be calculated from the MSD assuming Brownian like motion, where for long times the MSD has an asymptotic time dependence ${\rm MSD} \propto t$:
\begin{equation}\label{eq:diff}
D = \lim_{t\rightarrow\infty}\frac{{\rm MSD}}{4\,t}.
\end{equation}
However, in performing real experiments or numerical simulations, the systems of interest may behave slightly differently from the Brownian motion model and can have nonlinear time dependences, for example ${\rm MSD} \propto t^\alpha$, where the exponent $\alpha$ is a dimensionless parameter usually with a value close to unity. The case when $\alpha>1$ is called superdiffusion, and the opposite case  $\alpha<1$ is called subdiffusion. In both cases Eq.~(\ref{eq:diff}) is inconclusive and it is not possible to characterize the particle transport with a single parameter. Alternatively, it is possible to extend the concept of the diffusion coefficient to two parameters, namely the exponent $\alpha$ introduced previously and the generalized diffusion coefficient:
\begin{equation}\label{eq:diffa}
D_{\alpha} = \lim_{t\rightarrow\infty}\frac{{\rm MSD}}{4\,t^{\alpha}},
\end{equation}
as discussed in \cite{Feng14}.

The topic of anomalous diffusion in general is of high interest in various fields in physics and biology \cite{Anomal2018b}, where particle simulation methods provide significant contributions to the quantification of particle transport \cite{Anomal2018} because a solid theoretical background is still not available, especially in low dimensions \cite{YUAN19781}. Some predictions suggest that in the real thermodynamic limit (infinite system size and observation time) in isotropic systems with short range inter-particle interactions, the diffusion becomes normal and the instantaneous value of the diffusion exponent $\alpha$ asymptotically approaches unity \cite{Ott2009PRL}. However, for finite sizes and short times, highly relevant for nanotechnology and high frequency applications, the system can show significant anomalous transport, which can even be enhanced by the external magnetic field \cite{Hou09,KongWei2013,OttDiffu,Ott14}.  

Generally, as $t \rightarrow \infty$ is not directly accessible in particle simulations or experiments, a common practice is to substitute formulae (\ref{eq:diff}) and (\ref{eq:diffa}) by fitting the MSD curve  with ${\rm MSD}(t) = 4D_{\alpha} t^{\alpha} + b$ in a finite interval $t_1 < t < t_2$. In practice $t_1$ has to be chosen large enough for the time interval not to include the initial ballistic regime and possible oscillatory transients at early times \cite{Ott2009PRL}. The maximum time used in the fitting $t_2$ is determined by the physical size of the experiment, as the MSD is limited by the system size. 

It is essential that the results be presented in a form which allows their universal application, as well as allowing them to be compared with results from theory and numerical simulation. To do so we have to derive the principal parameters of the 2D magnetized YOCP model from our experiments. These parameters are:
\begin{itemize}
\item{the Coulomb coupling parameter
\begin{equation}
\Gamma = \frac{Q^2}{4\pi\varepsilon_0}\frac{1}{a\,k_{\rm B}T},
\end{equation}
where $\varepsilon_0$ is the dielectric constant, $k_{\rm B}$ is the Boltzmann constant, $T$ is the kinetic temperature of the dust grains, and $a$ is the Wigner-Seitz radius defined as $a=\sqrt{1/\pi n}$ in 2D, with $n$ being the surface number density of the dust grains,}
\item{the Yukawa screening parameter
\begin{equation}
\kappa = \frac{a}{\lambda_{\rm D}},
\end{equation}
where $\lambda_{\rm D}$ is the Debye screening length, a property representing the polarizability of the gas discharge plasma, and}
\item{the magnetization parameter
\begin{equation}
\beta = \frac{\omega_c}{\omega_p},
\end{equation}
where $\omega_p$ is the nominal 2D plasma frequency defined as $\omega_p^2 = nQ^2/2\varepsilon_0 m a$, and $\omega_c$ is the cyclotron frequency. In the case of a real magnetic field the cyclotron frequency can be calculated as $\omega_c=QB/m$, but in the RotoDust case, where the magnetic Lorentz force is substituted by the Coriolis force in the equation of motion of the dust grains, the cyclotron frequency is equivalent to $\omega_c=2\Omega$.}
\end{itemize}

To obtain the desired system parameters we perform the following steps for each measurement:
\begin{enumerate}
\item A calibration image is taken with the same optical setup to match the pixel size with physical distances. Our resolution is in the range of approximately 100 pixels per millimetre. As each dust grain covers approximately 5 pixels, the positions are known with sub-pixel accuracy, with the uncertainty in the measured inter-particle distance an estimated 5\%.
\item The number of observed dust grains and the visual size of the dust cloud are used to calculate the surface density, and via this, the Wigner-Seitz radius $a$.
\item To obtain the electric charge $Q$ and the Debye screening length $\lambda_{\rm D}$ we follow the procedure introduced in \cite{RotoDust}. After recording the image sequence of the rotating cloud, the rotation is stopped and all but two particles are dropped from the dust cloud by rapidly switching the discharge off and on in a short time but keeping all discharge parameters unchanged. Three parameters are easily obtainable from the recorded image sequence of this two-particle system: the average distance between the two particles $\langle d\rangle$, the oscillation frequency of the center of mass $\omega_{\rm COM}$ and the oscillation frequency of the inter-particle distance $\omega_{d}$. These three parameters, together with the grain mass $m$, are the input parameters for the solution of Eqs.~(2) and (3) of Ref.~\cite{Bonitz06}:
\begin{eqnarray}
\label{eq:5}
\frac{Q^2}{4\pi\varepsilon_0} &=& \frac12 m \omega^2_\text{COM} \frac{\langle d\rangle^3 \lambda_{\rm D}}{\langle d\rangle + \lambda_{\rm D}} e^{\langle d\rangle/\lambda_{\rm D}},  \\ 
\omega^2_\text{d} &=& 3\omega^2_\text{COM}\frac{\langle d\rangle^2+3\langle d\rangle\lambda_{\rm D}+3\lambda^2_{\rm D}}{\lambda_{\rm D}(\langle d\rangle+\lambda_{\rm D})},   \nonumber
\end{eqnarray}
which are derived using the assumptions of a harmonic trap for the horizontal confinement in form of $V
 _\text{tr}(r) = \frac12 M \omega^2_\text{COM} r^2$ and Yukawa interaction between the particles with potential energy 
\begin{equation}
\Phi_\text{Y}(r)=\frac{Q^2}{4\pi\varepsilon_0} \frac{\exp(-r/\lambda_{\rm D})}{r}.
\end{equation}
The solution of Eqs.~\ref{eq:5} provides the charge $Q$ and the Debye screening length $\lambda_{\rm D}$.
\item To obtain the most accurate estimation for the Coulomb coupling parameter $\Gamma$, we compute the pair correlation function $g(r)$ from the experimental particle position data and compare peak amplitudes to numerical data of non-magnetized 2D YOCP results as investigated in great detail in \cite{Ott11}. The mapping of magnetized and non-magnetized equilibrium pair correlation functions is guaranteed by the Bohr -- van Leeuwen theorem \cite{Leeuwen}.
\end{enumerate}
 
After performing all these additional steps, the results for the MSD and the diffusion coefficient are available as functions of the universal dimensionless parameters $\Gamma$, $\kappa$, and $\beta$ and will be presented together with the numerical results in section~\ref{sec:res}.

\section{\label{sec:sim} MD simulations }

Our numerical simulations are directly motivated by previous numerical studies of diffusion in magnetized 3D Yukawa systems \cite{OttDiffu,Begum2016}, studies on the connection between caging and diffusion in 2D Yukawa systems \cite{Ranna16}, and studies that identify superdiffusion in 2D magnetized systems \cite{Feng14,Wang2017}. Earlier investigations of non-magnetized Yukawa systems \cite{Lin98,Nunomura06,Ott09,Hou09,Shahzad12,Daligault12} provide valuable references for the methodology, the possible presence of superdiffusion, and numerical data.

We apply the molecular dynamics (MD) simulation method to describe the motion of the particles governed by Newton's equations of motion, where the forces included are due to the inter-particle Yukawa potential and the external magnetic field. Gas drag and random Brownian kicks originating from the background gas are neglected, as for the experimental conditions listed in Tables~\ref{tab:BUD} and \ref{tab:TEX} the Epstein dust-neutral collision frequency ($\nu_{\rm dn} = 0.5 - 30~{\rm sec}^{-1}$) is below the dust plasma frequency ($\omega_{\rm dp} = 100 - 300~{\rm rad/sec}$). For the integration of the equation of motion that accounts for the presence of the magnetic field we use the method described in \cite{Spreiter99}. In the simulations $N=4000$ particles are released in a 2D square simulation box with periodic boundary conditions. The simulation time-step is chosen to be short enough to resolve single particle oscillations; numerical stability is verified by monitoring the total kinetic energy in the system. During the initial ``thermalization'' phase of the simulation, a velocity back-scaling technique is applied to achieve the desired system temperature. This phase is kept long enough that the system reaches its stationary state. With increasing magnetic field the necessary relaxation time can become longer, as discussed in \cite{Ott13}. In the second ``measurement'' phase, no thermalization is applied and the particles move freely in the force field governed by the pairwise Yukawa interaction and the external magnetic field. From the simulated particle trajectories we derive the pair distribution function $g(r)$ and the mean square displacement MSD, the two quantities which are the focus of this study. 

\begin{figure}[htbp]
\begin{center}
\includegraphics[width=0.9\columnwidth]{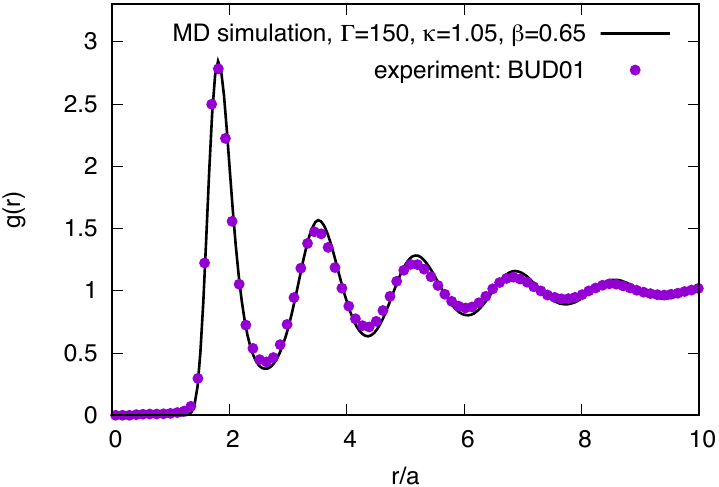}
\caption{Pair correlation functions $g(r)$ from the experiment (BUD01, symbols) and MD simulation (line) with parameters: $\Gamma=150$, $\kappa=1.05$, $\beta=0.65$.}
\label{fig:gr}
\end{center}
\end{figure}

Input parameters for the simulation, such as $\Gamma$, $\kappa$, and $\beta$ are taken from the experiments and are verified by comparing the experimental and computed pair correlation functions. In validating the measured and computed $g(r)$ data, the amplitude and position of the first peak is given the greatest weight, as long-term correlations are expected to be more affected by the finite size and confined geometry of the experimental system. We find agreement with deviations less than 10\% for the positions and amplitudes of the peak for the computed and measured $g(r)$ as demonstrated in figure~\ref{fig:gr}.

\section{\label{sec:res} Results }

Two series of measurements were carried out, one with the TEX setup and one with the BUD setup. Tables~\ref{tab:BUD} and \ref{tab:TEX} summarize the different cases and list the measured physical quantities.

\begin{table}[htp]
\caption{List of experiments on the BUD setup. Estimated uncertainties are within 1\% for $\Omega$, 5\% for pressure $p$, $a$, and $N$, and 10\% for $Q$, the RF Power $P_{\rm RF}$ and $\lambda_{\rm D}$. The dust particle diameter was $d_{\rm BUD} = 4.38 \,\mu{\rm m} \pm 1\%$.}
\begin{center}
\begin{tabular}{|c|c|c|c|c|c|c|c|c|}
\hline
exp. name & $p$  & $P_{\rm RF}$ & $\Omega$  & $Q$  & $a$ & $\lambda_{\rm D}$ & $N$  \\
 & [Pa] & [W] & [rad/s] & [$e$] & [$\mu$m] & [$\mu$m] & \\
\hline
BUD01 & 0.66 & 4.0  & 16.9 & 2330 & 235 & 230 & 158 \\
BUD02 & 1.0 & 1.0  & 18.2 & 2150 & 260 & 290 & 152\\
BUD03 & 1.05 & 1.0  & 23.2 & 2350 & 210 & 220 & 146\\
BUD04 & 1.05 & 3.0  & 23.1 & 2100 & 180 & 170 & 313\\ 
\hline
\end{tabular}
\end{center}
\label{tab:BUD}
\end{table}

\begin{table}[htp]
\caption{List of experiments on the TEX setup. Estimated uncertainties are within 1\% for $\Omega$, 5\% for pressure $p$, $a$, and $N$, and 10\% for $Q$, the RF Power $P_{\rm RF}$ and $\lambda_{\rm D}$. The dust particle diameter was $d_{\rm TEX} = 8.89 \,\mu{\rm m} \pm 1\%$}
\begin{center}
\begin{tabular}{|c|c|c|c|c|c|c|c|c|}
\hline
exp. name & $p$  & $P_{\rm RF}$ & $\Omega$  & $Q$  & $a$ & $\lambda_{\rm D}$ & $N$  \\
 & [Pa] & [W] & [rad/s] & [$e$] & [$\mu$m] & [$\mu$m] & \\
\hline
TEX01 & 12.0 & 4.2 & 11.09 & 3160 & 125 & 120 & 395 \\
TEX02 & 12.0 & 4.2 & 9.70 & 3160 & 108 & 120 & 395 \\
TEX03 & 12.0 & 4.2 & 8.31 & 3160 & 101 & 120 & 395 \\
TEX04 & 12.0 & 4.2 & 6.93 & 3160 & 97 & 120 & 395 \\
TEX05 & 13.3 & 5.1 & 11.11 & 3900 & 83 & 90 & 1910 \\
TEX06 & 13.3 & 5.1 & 11.09 & 3900 & 86 & 90 & 1910 \\
TEX07 & 8.0 & 4.3 & 19.25 & 2920 & 90 & 94 & 1720 \\
TEX08 & 13.3 & 6.4 & 55.66 & 5400 & 73 & 80 & 150 \\
TEX09 & 13.3 & 6.4 & 54.46 & 5400 & 73 & 80 & 150 \\
TEX10 & 13.3 & 6.4 & 54.97 & 5400 & 85 & 80 & 150 \\
TEX11 & 13.3 & 6.4 & 44.63 & 5400 & 61 & 80 & 150 \\
TEX12 & 26.6 & 6.1 & 41.16 & 4530 & 73 & 75 & 220 \\
TEX13 & 26.6 & 6.1 & 46.74 & 4530 & 69 & 75 & 220 \\
TEX14 & 20.0 & 8.2 & 40.79 & 5130 & 81 & 85 & 90 \\
TEX15 & 26.6 & 7.9 & 35.45 & 4620 & 76 & 85 & 100 \\
TEX16 & 20.0 & 6.7 & 28.52 & 5480 & 97 & 100 & 1200 \\
TEX17 & 20.0 & 6.7 & 13.89 & 5480 & 102 & 100 & 540 \\
TEX18 & 20.0 & 6.7 & 20.24 & 5480 & 111 & 100 & 390 \\
\hline
\end{tabular}
\end{center}
\label{tab:TEX}
\end{table}%

\begin{figure}[htbp]
\begin{center}
\includegraphics[width=0.9\columnwidth]{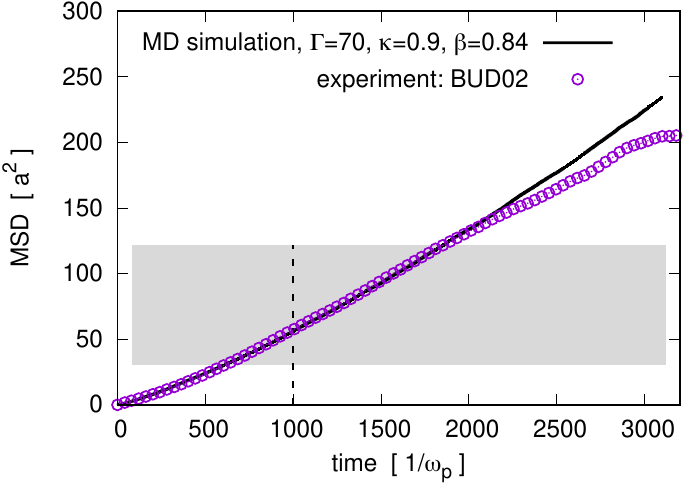}
\caption{MSD mean square displacement from the experiment (BUD02, symbols) and MD simulation (line) with parameters: $\Gamma=70$, $\kappa=0.9$, $\beta=0.84$. Only every 100th experimental data point is plotted for better visibility. The shaded area indicates the range for the parameter fitting, where the finite size effect is negligible. The shaded area itself does depend on the system size, as described in the text. The dashed line marks the time instance $t=1000/\omega_{\rm p}$ used for further analysis.}
\label{fig:msd}
\end{center}
\end{figure}

To illustrate the quality of the experimental MSD data an example is shown in figure~\ref{fig:msd}. The time interval shown here is much longer than the ballistic regime, which has a duration of approximately a few plasma oscillation cycles. The observed non-linearity is a true long time feature of the transport process, as supported by the numerical simulation. 

At large times ($t \omega_p > 2000$), or even more relevantly at large distances (${\rm MSD } > N a^2$, where $N$ is the number of particles in the dust cloud as listed in Tables~\ref{tab:BUD} and \ref{tab:TEX}), the trend of the experimental data changes, tending towards saturation, which is clearly a consequence of the final system size. This natural limitation means that these experiments can only be used to estimate the transport parameters which are based on the finite time characteristics of the MSD curve, as previously mentioned in Section~\ref{sec:exp}.

To obtain the quantities $\alpha$ and $D_\alpha$ that are used to characterize the anomalous diffusion following the definition in Eq.~(\ref{eq:diffa}) we perform least square fitting to both the simulation and experimental data in the form ${\rm MSD}(t) = 4D_\alpha t^\alpha + b$. The fitting is performed in the parameter range $0.2N < {\rm MSD}/a^2 < 0.8N$, to minimize the effects of the  finite size saturation.

The dimensionless YOCP parameters $\Gamma$, $\kappa$, and $\beta$ derived for each experimental condition are listed In Table~\ref{tab:3}.  A comparison is  given for the anomalous diffusion parameters $\alpha$ and $D_{\alpha}$ derived from the experiment and MD simulation for each case.

\begin{table}[htp]
\caption{Dimensionless system parameters and diffusion coefficients. Estimated uncertainties are within 15\% for all quantities. The unit of $D_{\alpha}$ is $100 a^2/\omega_{\rm p}^{\alpha}$.}
\begin{center}
\begin{tabular}{|c|c|c|c|c|c|c|c|c|}
\hline
exp. & $\Gamma$ & $\kappa$ & $\beta$ & $\alpha$ & $D_{\alpha}$ &  $\alpha$ & $D_{\alpha}$ \\
name   &  &  &  & (exp.) & (exp.) &  (MD) & (MD) \\
\hline
BUD01 & 45  & 1.0 & 0.63 & 1.07 & 1.28 & 1.10 & 1.19  \\
BUD02 & 70  & 0.9 & 0.84 & 1.26 & 0.182 & 1.25 & 0.197 \\
BUD03 & 170  & 1.0 & 0.77 & 1.16 & 0.075 & 1.10 & 0.097 \\
BUD04 & 150  & 1.05 & 0.65 & 1.24 & 0.072 & 1.30 & 0.063 \\
TEX01 & 61 & 1.0 & 0.46 & 1.07 & 1.01 & 1.11 & 1.01 \\
TEX02 & 63 & 0.9 & 0.37 & 1.14 & 0.800 & 1.17 & 0.558 \\
TEX03 & 66 & 0.85 & 0.30 & 1.22 & 0.402 & 1.27 & 0.304 \\
TEX04 & 70 & 0.8 & 0.24 & 1.36 & 0.150 & 1.35 & 0.168 \\
TEX05 & 45 & 0.9 & 0.23 & 1.39 & 0.229 & 1.14 & 1.37 \\
TEX06 & 78 & 0.93 & 0.24 & 1.20 & 0.360 & 1.14 & 0.659 \\
TEX07 & 24 & 0.96 & 0.55 & 1.12 & 2.15 & 1.04 & 4.10 \\
TEX08 & 135 & 0.9 & 0.69 & 1.17 & 0.114 & 1.18 & 0.107 \\
TEX09 & 135 & 0.9 & 0.71 & 1.18 & 0.116 & 1.23 & 0.073 \\
TEX10 & 122 & 1.0 & 0.89 & 1.06 & 0.351 & 1.10 & 0.222 \\
TEX11 & 138 & 0.85 & 0.50 & 1.00 & 0.361 & 1.05 & 0.270 \\
TEX12 & 115 & 1.0 & 0.64 & 1.21 & 0.116 & 1.24 & 0.106 \\
TEX13 & 58 & 0.9 & 0.61 & 1.19 & 0.434 & 1.17 & 0.537 \\
TEX14 & 53 & 0.95 & 0.61 & 1.08 & 1.02 & 1.13 & 0.804 \\
TEX15 & 41 & 0.9 & 0.45 & 1.01 & 3.07 & 1.04 & 2.60 \\
TEX16 & 81 & 1.0 & 0.52 & 1.17 & 0.328 & 1.12 & 0.619 \\
TEX17 & 114 & 1.0 & 0.35 & 1.16 & 0.281 & 1.16 & 0.267 \\
TEX18 & 71 & 0.9 & 0.41 & 1.00 & 1.63 & 1.09 & 0.818 \\
\hline
\end{tabular}
\end{center}
\label{tab:3}
\end{table}%

The anomalous diffusion exponent $\alpha$ takes on values in the range between 1.0 and 1.4, a consequence of superdiffusive behaviour. The values of the generalized diffusion coefficient $D_{\alpha}$ depend sensitively on the exponent $\alpha$, making direct comparison of experimental and simulation results difficult. Therefore, we define a fixed-time diffusion coefficient $D^{1000}$, which is calculated as
\begin{equation}
    D^{1000} = D(t=1000/\omega_{\rm p}) = \frac{{\rm MSD}(t=1000/\omega_{\rm p})}{4 \times 1000/\omega_{\rm p}}.
\end{equation}

Being a finite-time quantity $D^{1000}$ may not properly represent the particle transport in the thermodynamic limit, but similar concepts could be useful in applications related to small samples (nanotechnology) and ultra-fast processes, where spatial or temporal constraints are present. $D^{1000}$ is still a quantity that strongly depends on all relevant system parameter ($\Gamma$, $\kappa$, and $\beta$); however it has been shown that the relative diffusion coefficient $D/D_0$, the ratio of the magnetized ($\beta > 0$) and the non-magnetized ($\beta = 0$) values for given $\Gamma$ and $\kappa$ parameters, is a function of the magnetization $\beta$ only \cite{Ott14} as
\begin{equation}\label{eq:Ott14}
    \frac{D}{D_0}(\beta)=\frac{1+\frac{1}{3}\beta}{1+\frac{7}{4}\beta+\beta^2}.
\end{equation}

\begin{figure}[htbp]
\begin{center}
\includegraphics[width=0.9\columnwidth]{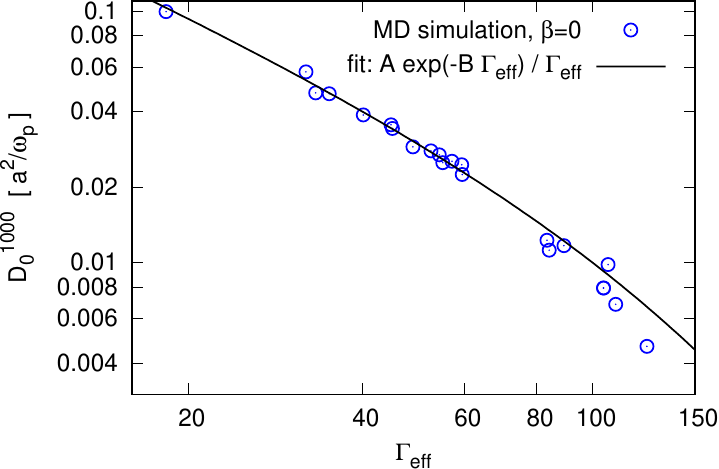}
\caption{Fixed-time ($t\omega_{\rm p} = 1000$) diffusion coefficient, $D_0^{1000}$, from the $\beta = 0$ reference simulations as a function of the effective Coulomb coupling parameter. The theoretical fit formula is adapted from Ref.~\cite{Daligault12} with $A=2.1444$ and $B=0.00778$.}
\label{fig:D0}
\end{center}
\end{figure}

To derive the relative diffusion coefficients, we first computed the MSD and $D^{1000}_0$ values for the non-magnetized ($\beta=0$) 2D Yukawa systems with Yukawa parameters $\Gamma$ and $\kappa$ corresponding to the experimental cases listed in Table~\ref{tab:3}. The resulting values are plotted in Fig.~\ref{fig:D0} and listed in Table~\ref{tab:4}. From the Yukawa system parameters $\Gamma$ and $\kappa$ we derived the effective Coulomb coupling coefficient $\Gamma_{\rm eff}$ based on the height of the first maximum of $g(r)$ as defined for the liquid regime in \cite{Ott11}. Unfortunately in the experiment $\Gamma$ and $\kappa$ can not be set arbitrarily, and we can not attain a one-to-one match for the conditions in the magnetized and unmagnetized cases. In the MD simulations, however, these are the main input parameters, therefore the simulation values of $D^{1000}_0$ are used to derive $D^{1000}/D^{1000}_0$ for both the simulation and the experiment. 

As shown in figure~\ref{fig:D0}, the reference values $\beta=0$ for the fixed-time diffusion coefficients show a dependence on the Coulomb coupling coefficient, which is well approximated by the formula
\begin{equation}
    \frac{D^{1000}_0}{a^2\omega_{\rm p}} = \frac{A(\kappa)}{\Gamma}\exp\left[-B(\kappa)\Gamma\right]
\end{equation}
derived for 3D Yukawa systems in \cite{Daligault12}, with somewhat different numerical factors $A(\kappa)=A=2.1444$ and $B(\kappa)=B=0.00778$, where the $\kappa$-dependent coefficients are approximated by constants, as $\kappa$ shows only small variations around $\kappa=1$.

\begin{table}[htp]
\caption{Dimensionless system parameters and diffusion coefficients. Estimated uncertainties are within 15\% for all quantities. The unit of $D^{1000}$ is $100 a^2/\omega_{\rm p}$.}
\begin{center}
\begin{tabular}{|c|c|c|c|c|c|c|c|c|}
\hline
exp. & $\Gamma_{\rm eff}$ & $\beta$ & $D^{1000}_0$ & $D^{1000}$ & $D^{1000}$  \\
name & & & (MD) & (MD) & (exp.) \\
\hline
BUD01 & 33.2 & 0.63 & 4.74 & 2.32 & 2.02 \\
BUD02 & 54.3 & 0.84 & 2.69 & 1.10 & 1.08 \\
BUD03 & 124 & 0.77 & 0.466 & 0.189 & 0.221 \\
BUD04 & 106 & 0.65 & 0.985 & 0.487 & 0.369 \\
TEX01 & 44.8 & 0.46 & 3.54 & 2.09 & 1.59 \\
TEX02 & 48.9 & 0.37 & 2.89 & 1.79 & 2.09 \\
TEX03 & 52.5 & 0.30 & 2.79 & 1.94 & 1.82 \\
TEX04 & 57.1 & 0.24 & 2.54 & 1.85 & 1.76 \\
TEX05 & 35.0 & 0.23 & 4.70 & 3.57 & 3.36 \\
TEX06 & 59.5 & 0.24 & 2.25 & 1.70 & 1.41 \\
TEX07 & 18.3 & 0.55 & 9.96 & 5.33 & 4.85 \\
TEX08 & 104.3 & 0.69 & 0.795 & 0.362 & 0.363 \\
TEX09 & 104.3 & 0.71 & 0.795 & 0.352 & 0.398 \\
TEX10 & 89.1 & 0.89 & 1.17 & 0.435 & 0.523 \\
TEX11 & 109.5 & 0.50 & 0.685 & 0.371 & 0.351 \\
TEX12 & 84.1 & 0.64 & 1.12 & 0.541 & 0.480 \\
TEX13 & 45.0 & 0.61 & 3.42 & 1.70 & 1.58 \\
TEX14 & 40.1 & 0.61 & 3.87 & 1.93 & 1.74 \\
TEX15 & 31.9 & 0.45 & 5.75 & 3.42 & 3.28 \\
TEX16 & 59.4 & 0.52 & 2.46 & 1.39 & 1.04 \\
TEX17 & 83.3 & 0.35 & 1.23 & 0.805 & 0.850 \\
TEX18 & 55.0 & 0.41 & 2.50 & 1.51 & 1.62 \\
\hline
\end{tabular}
\end{center}
\label{tab:4}
\end{table}

\begin{figure}[b]
\begin{center}
\includegraphics[width=0.9\columnwidth]{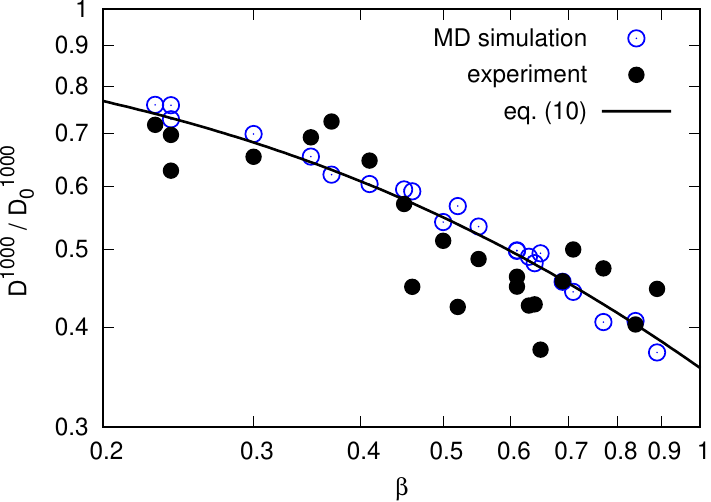}
\caption{Fixed-time ($t\omega_{\rm p} = 1000$) diffusion coefficient relative to the non-magnetized values, $D_0$, as a function of the dimensionless magnetization parameter $\beta$ parameter.}
\label{fig:DpD0}
\end{center}
\end{figure}

The final results of this study are listed in table~\ref{tab:4}, where numerical values of the fixed-time diffusion coefficients $D^{1000}$ from both the dusty plasma experiments and the corresponding MD simulations are given. Graphical representation of the data is shown in figure~\ref{fig:DpD0}, where the relative diffusion coefficients $D^{1000}/D^{1000}_0$ are plotted for both the experiments and the simulations as a function of the magnetization parameter $\beta$. The analytic formula given by eq.~(\ref{eq:Ott14}) is shown as a line which is a good representation of the data points. The simulation data closely follows the theoretical trend. The experimental results have significantly higher scatter and uncertainty, but generally support the model prediction.

\section{\label{sec:sum} Summary}

RotoDust experiments and molecular dynamics simulations were carried out to quantify the diffusion (mass transport) in quasi-magnetized single layer dusty plasmas and to link this to existing transport model results for magnetized two-dimensional Yukawa systems. Although the relatively small size of the experimental dust cloud limits our investigations to the range of time far from the thermodynamic limit and the Yukawa system parameters cannot be controlled independently as in simulations, we were able to confirm two of the main theoretical predictions for strongly coupled magnetized 2D Yukawa plasmas~\cite{Ott14,Feng14}.

(i) At intermediate times, the experimental MSD curves of quasi-magnetized systems clearly show superdiffusive behavior, where the MSD grows faster than linear with time. The exponents $\alpha$ strongly depend on the coupling, screening, and effective magnetization parameter and are in good agreement with results from molecular dynamics simulations. (ii) The relative fixed-time diffusion coefficient, which characterizes mass transport on intermediate time scales, was shown to be consistent with a scaling-law~\cite{Ott14} that is largely independent of the screening and coupling parameters. It describes the decrease of the particles' mobility with an increase of the (effective) magnetization. In our experiments, the (relative) mobility was reduced by a factor $\sim 2$ at the highest effective magnetization, $\beta\sim 0.9$.

\begin{acknowledgments}
The authors are grateful for financial support from the Hungarian Office for Research, Development, and Innovation NKFIH grants K-119357 and K-115805, the J\'anos Bolyai Research Scholarship of the Hungarian Academy of Sciences, the National Science Foundation grants PHY-1414523 and PHY-1707215 and the program BR 05236730 of the Ministry of Education and Science of the Republic of Kazakhstan.
\end{acknowledgments}

%

\end{document}